\documentclass[12pt,titlepage]{article}
\usepackage{amsmath}
\usepackage{amssymb}
\usepackage{graphicx}
\usepackage{caption2}
\usepackage{amsfonts}
\usepackage{cite}

\newcommand{\be}{\begin{equation}}
\newcommand{\ee}{\end{equation}}
\newcommand{\bea}{\begin{eqnarray}}
\newcommand{\eea}{\end{eqnarray}}
\newcommand{\bef}{\begin{figure}}
\newcommand{\ef}{\end{figure}}
\newcommand{\bt}{\begin{tabular}}
\newcommand{\et}{\end{tabular}}
\newcommand{\bno}{\begin{enumerate}}
\newcommand{\eno}{\end{enumerate}}

\begin{document}

\begin{center}

{\bf\large  On peaked solitary waves of the Degasperis - Procesi  equation}

\vspace{0.5cm}

Shijun Liao\\
\vspace{0.5cm}
State Key Laboratory of Ocean Engineering, Dept. of Mathematics\\
School of Naval Architecture, Ocean and Civil Engineering\\
Shanghai Jiao Tong University, Shanghai 200240, China\\
\vspace{0.5cm}
(Email address:  sjliao@sjtu.edu.cn)

\end{center}

\hspace{-0.75cm} {\bf Abstract} {\em  The  Degasperis - Procesi  (DP) equation  describing  the propagation of  shallow water waves contains a physical parameter $\omega$, and it is well-known that  the DP equation admits  solitary waves with a peaked crest when $\omega = 0$.    In this article, we illustrate, for the first time,  that the DP equation admits peaked solitary waves even when $\omega \neq 0$.  This is helpful to enrich our knowledge and deepen our understandings about peaked solitary waves of the DP equation.       }

\hspace{-0.75cm} {\bf PACS Number}: 47.35.Bb

\hspace{-0.75cm} {\bf Key Words}  Peaked solitary waves,  discontinuity, Degasperis - Procesi  equation

\section{Introduction}

The solitary surface wave was first discovered by  Russell \cite{Russell1844} in 1834.  Since then,  many models for shallow water waves  have been developed, such as the Boussinesq equation \cite{Boussinesq1872}, the  Korteweg  \&  de Vries (KdV)  equation \cite{KdV} and  the  Benjamin-Bona-Mahony   (BBM) equation \cite{Benjamin1972},  and so on.  All of them admit solitary waves with smooth crest.  
   In contrast to the KdV, Boussinesq and BBM equations,   the celebrated Camassa-Holm (CH) equation \cite{Camassa1993PRL}
\begin{equation}
u_t+2\omega u_x-u_{xxt}+ 3 u u_x = 2 u_x u_{xx} + u u_{xxx}\label{geq:CH}
\end{equation}
can model  both  phenomena of soliton interaction and wave breaking (see \cite{Constantin2000}),  where $u(x,t)$ denotes the wave elevation, $x,t$ are the temporal and spatial variables, $\omega$ is a constant related to the critical shallow water wave speed,  and the subscript denotes the partial differentiation, respectively.    The CH equation (\ref{geq:CH}) is integrable and bi-Hamiltonian.    Especially, when $\omega=0$,   the CH equation (\ref{geq:CH}) has the peaked solitary wave \cite{Camassa1993PRL}
\[  u(x,t)  =  c \exp (-|x-c t|). \]

Degasperis  and Procesi  \cite{DP1999} investigated a generalized equation 
\begin{equation}
u_t+2\omega u_x-u_{xxt}+ (\beta + 1) u u_x = \beta u_x u_{xx} + u u_{xxx}\label{geq:generalized:1}
\end{equation}
and found that it is completely integrable only when $\beta=2$, corresponding to the CH equation (\ref{geq:CH}),  and $\beta = 3$, corresponding to the so-called Degasperis - Procesi  (DP) equation     
\begin{equation}
u_t+2\omega u_x-u_{xxt}+ 4 u u_x = 3 u_x u_{xx} + u u_{xxx},  \label{geq:DP}
\end{equation}
respectively.  The DP equation (\ref{geq:DP}) can model the propagation of shallow water waves with small amplitude and large wave-length.    It  is interesting that,  like the CH equation (\ref{geq:CH}),  the DP equation (\ref{geq:DP}) also admits  solitary waves with a peaked crest when $\omega=0$.    

Currently,   Liao \cite{Liao-arXiv-KdV}  gained the closed-form solutions of the peaked solitary waves of the KdV equation, the modified KdV equation, the Boussinesq equation and the BBM equation.  Besides,
by means of the Mathematica package BVPh 1.0 that is based on the homotopy analysis method (HAM) \cite{Liao1999, Liao1999JFM, LiaoBook2003, LiaoBook2012},  a powerful analytic method for highly nonlinear  equations,  
 Liao \cite{Liao-arXiv-CH}  found that  the CH equation (\ref{geq:CH})    also admits  peaked solitary waves  even in the case of $\omega \neq 0$.   All of these indicate that most of mainstream models of shallow water waves  admit  peaked solitary waves.   

In this article, using the HAM-based Mathematica package BVPh 1.0 for nonlinear boundary value problems in a similar way,  we illustrate that the DP equation (\ref{geq:DP}) also admits peaked solitary waves even when $\omega\neq 0$.   In addition,  we  further  illustrate that  Eq. (\ref{geq:generalized:1})  for arbitrary constant $\beta$ and  the more generalized equation   
\begin{equation}
u_t+2\omega u_x-u_{xxt}+ \alpha  u u_x = \beta u_x u_{xx} + u u_{xxx}\label{geq:generalized:2}
\end{equation}
for arbitrary constants $\alpha$ and $\beta$ also  admit peaked solitary waves even when $\omega\neq 0$, although they are not  integrable in general cases.  

\section{Peaked solitary waves of  Eq.  (\ref{geq:generalized:2}) in the case of $\omega \neq 0$ }

Let us first consider the propagating solitary waves of the DP equation   (\ref{geq:DP})  with permanent form, corresponding to $\alpha=4$ and $\beta = 3$ of Eq. (\ref{geq:generalized:2}).     To avoid the repeat,   Eq. (\ref{geq:generalized:2}) is used to describe the basic ideas of our analytic approach.      Writing $\xi = x- c \; t$ and $ w(\xi) = c  \;  u(x,t) $,   the original equation  (\ref{geq:generalized:2})   becomes
\begin{equation}
w''' - \left(1-\frac{2\omega}{c}\right)w' + \alpha  w w' = \beta  w' w'' + w w''',  \label{geq:generalized:3}
\end{equation}
subject to the boundary conditions
\begin{equation}
w\to 0, w'\to 0, w''\to 0, \;\; \mbox{as $|\xi|\to +\infty$}, \label{bc:generalized:3}
\end{equation}
where the prime denotes the differentiation.  

The  corresponding  linearized equation
\begin{equation}
w''' - \left(1-\frac{2\omega}{c}\right) w' =0 \label{geq:linear}
\end{equation}
has the peaked solitary wave
\begin{equation}
w(\xi) = A\; \exp(-\mu |\xi|),  \;\;  \mu = \sqrt{1-\frac{2\omega}{c}},
\end{equation}
provided $\omega < c/2$, where $A = w(0)$.    The solution of Eq.  (\ref{geq:generalized:3})  can be expressed in the form
\[  w(\xi) = \sum_{n=1}^{+\infty} a_n \; \exp(-n \mu |\xi|), \]
where $a_n$ is a consant to be determined.    Using the homotopy analysis method (HAM) \cite{Liao1999, Liao1999JFM, LiaoBook2003, LiaoBook2012},  an analytic technique for highly nonlinear differential equations, we  gain the series solution
\begin{equation}
w(\xi) = w_0(\xi) +\sum_{m=0}^{+\infty} w_m(\xi).\label{series:w}
\end{equation}
Here
\[  w_0(\xi) = A \exp(-\mu |\xi|) \]
is the initial guess,  $w_m$ for $m\geq 1$ is governed by
\begin{equation}
{\cal L}\left[ w_m(\xi) -\chi_m \; w_{m-1}(\xi)  \right] = c_0 \; \delta_{m-1}(\xi),
\end{equation}
subject to the boundary condition
\begin{equation}
w_m(0)=0, \;  w_m\to 0, \;\;  \mbox{as $|\xi|\to +\infty$},
\end{equation}
where 
\[  {\cal L}f = f''' -\mu^2 f' \]
is an auxiliary linear operator, and
\begin{eqnarray}
\delta_n &=&  w_n'''-\mu^2 w_n' + \sum_{j=0}^{n} \left[ \beta  \;  w_j  \; w'_{n-j} -\alpha  \;  w'_j  \; w''_{n-j} - w_j \; w'''_{n-j} \right],\;\;\;\;\\
\chi_n &=& \left\{
\begin{array}{ll}
1 & \mbox{when $n>1$},\\
0 & \mbox{otherwise.}
\end{array}
\right.
\end{eqnarray}
Note that $c_0\neq 0 $ is an auxiliary parameter, called the convergence-control parameter,  
which  provides us a convenient way to guarantee the convergence of the approximation series.   For details, please refer to Liao \cite{Liao1999, LiaoBook2003, LiaoBook2012}.  In fact,  directly using the HAM-based Mathematica package BVPh 1.0 (see Part II of  Liao's book \cite{LiaoBook2012}) for nonlinear boundary-value/eigenvalue problems,  it is straightforward to gain high-order  analytic approximations of   (\ref{geq:generalized:3}) and (\ref{bc:generalized:3}).    For details, please refer to the Appendix.

\begin{table}[t]
\begin{center}
\caption{$u'(0_+)$ of the $m$th-order analytic approximations and the  residual squares of the DP equation (\ref{geq:DP}) in  the  case of $c=1, \omega=1/4$, and $A = \pm 1/10$ given by the HAM-based Mathematica package BVPh 1.0  with the convergence-control parameter $c_0=-1$.  }

\vspace{0.25cm}

\begin{tabular}{|c|c|c|c|c|}
  \hline\hline
   &  \multicolumn{2}{c|}{$A=1/10$}  & \multicolumn{2}{c|}{$A=-1/10$}  \\ \cline{2-5}
  $m$ &  $u'(0_+)$ & ${\cal E}_m$ & $u'(0_+)$ & ${\cal E}_m$ \\ \hline
  2 & -0.065250 & 3.8 E-9 & 0.074675 & 4.8E-9 \\
  \hline
  4 & -0.065110 & 2.0 E-12 & 0.074779 & 1.6E-12 \\
  \hline
  6 & -0.065105 & 9.2 E-15 & 0.074782 & 5.9E-15 \\
  \hline
  8 & -0.065105 & 2.4 E-17 & 0.074782 & 1.6E-17 \\
  \hline
  10 & -0.065105 & 5.2 E-20 &0.074782 & 3.5E-20 \\
  \hline\hline
\end{tabular}
\label{Table:DP}
\end{center}
\end{table}

 \begin{figure}[thbp]
\centering
\includegraphics[scale=0.4]{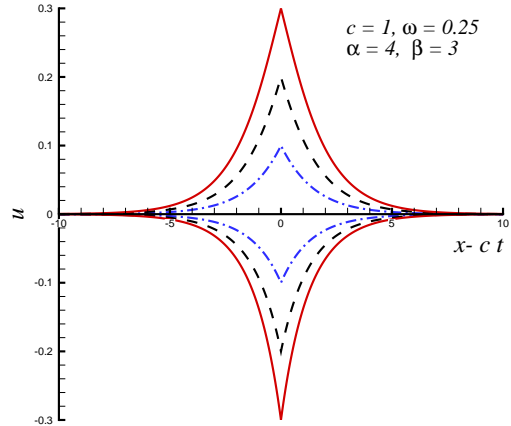}
\caption{ The peaked solitary waves $u(x,t)$ of  the DP equation (\ref{geq:DP}) in the case of $\omega=1/4$ and $c = 1$.  Solid line: $A=\pm 3/10$;  Dashed line: $A = \pm 1/5$;  Dash-dotted line: $A = \pm 1/10$. }
\label{figure:DP}
\end{figure}

 The accuracy of the $m$th-order approximation of $w(\xi)$  is defined by the averaged residual square
of the governing equation (\ref{geq:generalized:3}) in the domain $\xi\in[0,a]$:
\begin{equation}
{\cal E}_m = \frac{1}{a} \int_0^a \left[{\cal N}\left(\sum_{j=0}^{m}w_j\right) \right]^2 d\xi,
\end{equation}
where
\[ {\cal N}w = w''' - \mu^2 w' + \alpha \;  w w' - \beta \;  w' w'' - w w'''. \]
We choose  $a=10$ in this article, because the wave elevation decays exponentially.  

For simplicity, we  investigate  the case $c = 1$ in this article.   Without loss of generality,  we  first  consider  the case of $\omega = 1/4$ and  $A = \pm 1/10$ for the DP equation (\ref{geq:DP}).   As shown in Table~\ref{Table:DP}, the averaged residual squares  of  the 10th-order analytic approximations decrease to 5.2$\times 10^{-20}$  in the case of $A=1/10$ and to 3.5$\times 10^{-20}$ in the case of $A=-1/10$, respectively.    Besides,  the corresponding values of $u'(0_+)$ (the limit is taken as $\xi \to 0$ from the right) quickly  converge to -0.065105 and 0.074782, respectively.    All of these clearly indicate that the series given by  the Mathematica package BVPh 1.0  converge to the solution of the DP equation (\ref{geq:DP}).   The solutions for larger values of  $|A|$ can be gained in a similar way.  All of these solitary waves have a peaked crest, as shown in Fig.~\ref{figure:DP}.   These clearly illustrate  that, like the CH equation (\ref{geq:CH}),   the DP equation (\ref{geq:DP}) also admits peaked solitary waves even when $\omega \neq 0$.   Besides,  like the CH equation (\ref{geq:CH}),  the phase speed of the peaked solitary waves of the DP equation (\ref{geq:DP})  has nothing to do with the wave amplitude $A$.  This is quite interesting.

As pointed out by Degasperis  and Procesi  \cite{DP1999}, the generalized equation (\ref{geq:generalized:1}) is completely integrable only when $\beta=2$ or $\beta=3$, corresponding to the CH equation (\ref{geq:CH}) or the DP equation (\ref{geq:DP}), respectively.    Does Eq.  (\ref{geq:generalized:1}) admit  peaked solitary waves when $\beta\neq 2$ and $\beta\neq 3$ ?

\begin{table}[t]
\begin{center}
\caption{$u'(0_+)$ of the $m$th-order analytic approximations and the  residual squares of Eq.  (\ref{geq:generalized:1}) in the case of $c=1, \omega=1/4, A = \pm 1/10$ and  $\beta = 5/2$ given by the HAM-based Mathematica package BVPh 1.0 with the convergence-control parameter $c_0=-1$.  }

\vspace{0.25cm}

\begin{tabular}{|c|c|c|c|c|}
  \hline\hline
   &  \multicolumn{2}{c|}{$A= 1/10$}  & \multicolumn{2}{c|}{$A=-1/10$}  \\ \cline{2-5}
  $m$ &  $u'(0_+)$ & ${\cal E}_m$ & $u'(0_+)$ & ${\cal E}_m$ \\ \hline
  2 & -0.066002 & 2.1 E-9   & 0.074251 & 2.6E-9 \\   \hline
  4 & -0.065904 & 5.7 E-13 & 0.074323 & 5.3E-13 \\ \hline
  6 & -0.065901 & 1.6 E-15 & 0.074325 & 1.1E-15 \\ \hline
  8 & -0.065901 & 3.0 E-18 & 0.074325 & 2.2E-18 \\ \hline
10 & -0.065901 & 4.4 E-21 & 0.074325 & 3.2E-21 \\
  \hline\hline
\end{tabular}
\label{Table:generalized:1}
\end{center}
\end{table}

 \begin{figure}[thb]
\centering
\includegraphics[scale=0.4]{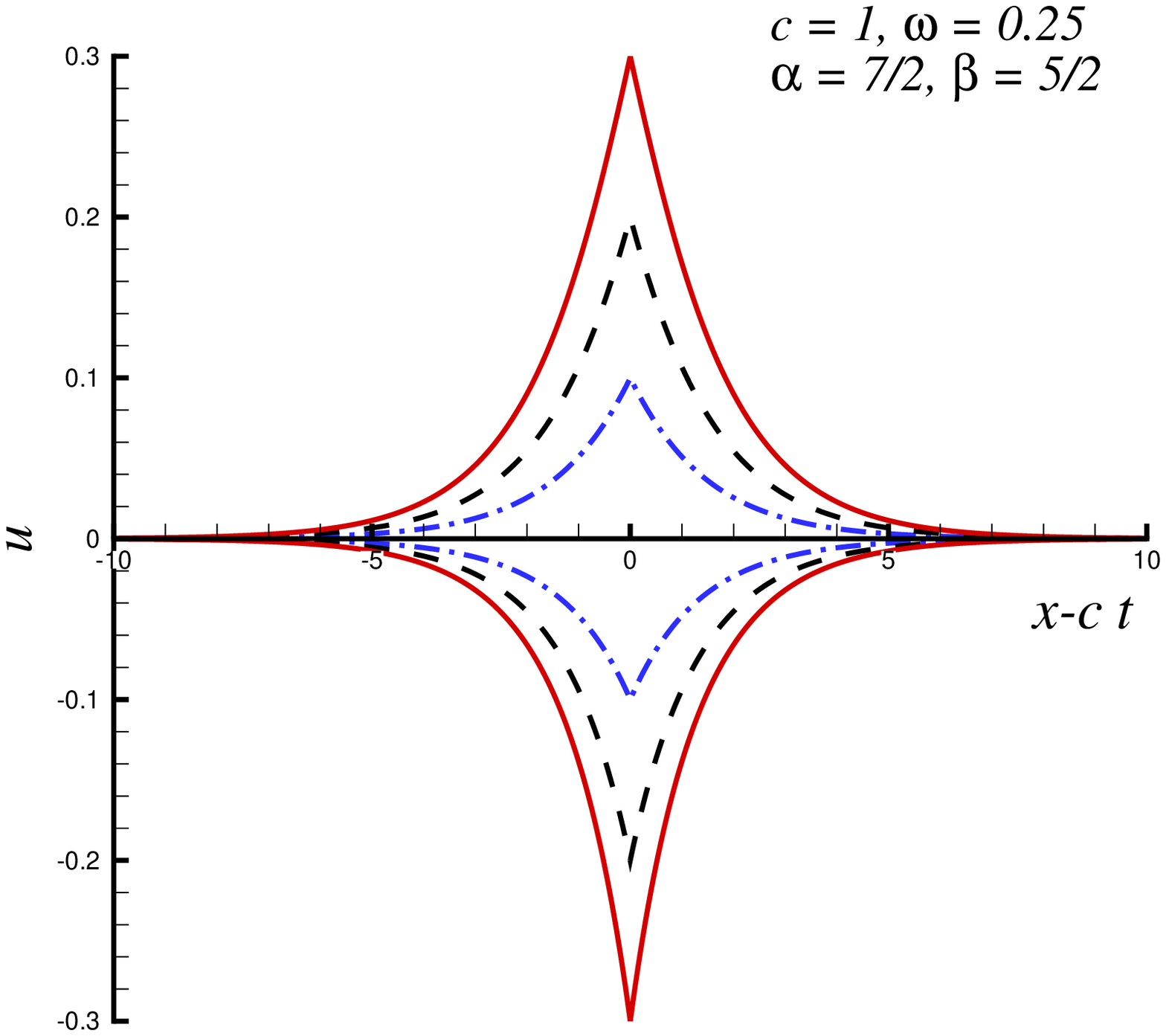}
\caption{ The peaked solitary waves $u(x,t)$ of  Eq. (\ref{geq:generalized:1}) in the case of $\omega=1/4$, $c = 1$ and  $\beta=5/2$, corresponding to $\alpha=7/2, \beta=5/2$ of Eq. (\ref{geq:generalized:2}).   Solid line: $A=\pm 3/10$;  Dashed line: $A = \pm 1/5$;  Dash-dotted line: $A = \pm 1/10$.  }
\label{figure:generalized:1}
\end{figure}

Without loss of generality,  let us consider the case of $c=1, \omega = 1/4, \beta = 5/2$ and $A = \pm 1/10$ for Eq. (\ref{geq:generalized:1}), corresponding to $\alpha=7/2, \beta=5/2$  for  Eq. (\ref{geq:generalized:2}).  Similarly, by means of the  HAM-based Mathematica package BVPh 1.0, we gain the convergent approximations with $c_0 = -1$.    As shown in Table~\ref{Table:generalized:1}, the averaged residual squares  of  the 10th-order analytic approximations  decrease  to 4.4$\times 10^{-21}$  in the case of $A=1/10$ and to 3.2$\times 10^{-21}$ in the case of $A=-1/10$, respectively.   Similarly,  using the Mathematica package BVPh 1.0,  we gain the convergent approximations for larger values of $A$, as shown in  Fig.~\ref{figure:generalized:1}.  Note that all of these solitary waves have a peaked crest!  Thus,  the generalized  equation (\ref{geq:generalized:1}) in the case of $\omega \neq 0$  also admits peaked solitary waves,  even when $\beta\neq 2$ and $\beta \neq 3$ so that it  is {\em not}  integrable!     

\begin{table}[t]
\begin{center}
\caption{$u'(0_+)$ of the $m$th-order analytic approximations and the  residual squares of the generalized  equation (\ref{geq:generalized:2}) in the case of $c=1, \omega=1/4, A = \pm 1/10,  \alpha = 11/3$ and  $\beta = 5/2$ given by the HAM-based Mathematica package BVPh 1.0 with the convergence-control parameter $c_0=-1$.  }

\vspace{0.25cm}

\begin{tabular}{|c|c|c|c|c|}
  \hline\hline
   &  \multicolumn{2}{c|}{$A= 1/10$}  & \multicolumn{2}{c|}{$A=-1/10$}  \\ \cline{2-5}
  $m$ &  $u'(0_+)$ & ${\cal E}_m$ & $u'(0_+)$ & ${\cal E}_m$ \\ \hline
  2 & -0.065541 & 2.3 E-9   & 0.074576 & 2.9E-9 \\   \hline
  4 & -0.065428 & 9.8 E-13 & 0.074658 & 7.6E-13 \\ \hline
  6 & -0.065425 & 3.3 E-15 & 0.074660 & 2.2E-15 \\ \hline
  8 & -0.065425 & 6.6 E-18 & 0.074660& 4.6E-18 \\ \hline
10 & -0.065425 & 1.1 E-20 & 0.074660 & 7.8E-21 \\
  \hline\hline
\end{tabular}
\label{Table:generalized:2}
\end{center}
\end{table}

 \begin{figure}[thbp]
\centering
\includegraphics[scale=0.4]{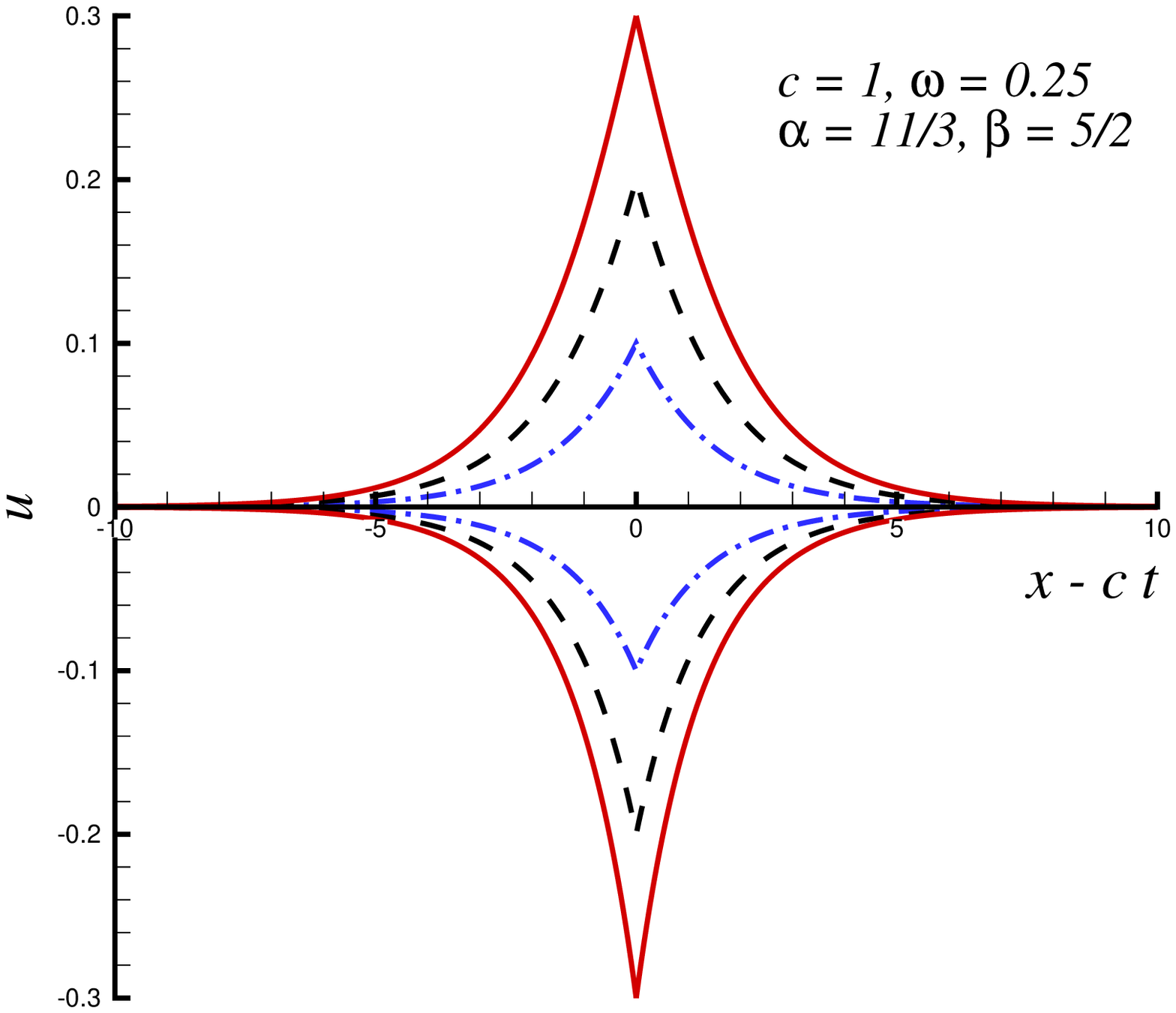}
\caption{ The peaked solitary waves $u(x,t)$ of  (\ref{geq:generalized:2}) in the case of $\omega=1/4, c = 1$, $\alpha=11/3$ and $\beta = 5/2$.   Solid line: $A=\pm 3/10$;  Dashed line: $A = \pm 1/5$;  Dash-dotted line: $A = \pm 1/10$. }
\label{figure:generalized:2}
\end{figure}

How about the more generalized equation (\ref{geq:generalized:2})?  

Without loss of generality, let us consider  the generalized equation (\ref{geq:generalized:2}) in the case of $c=1, \omega = 1/4, \alpha=11/3, \beta = 5/2$ and $A=\pm 1/10$.     Similarly, by means of the  HAM-based Mathematica package BVPh 1.0, we gain the convergent approximations with $c_0 = -1$.    As shown in Table~\ref{Table:generalized:2}, the averaged residual squares  of  the 10th-order analytic approximations  decrease to 1.1$\times 10^{-20}$  in the case of $A=1/10$ and to 7.8$\times 10^{-21}$ in the case of $A=-1/10$, respectively.   Similarly,  using the Mathematica package BVPh 1.0,  we gain the convergent approximations for larger values of $|A|$, as shown in  Fig.~\ref{figure:generalized:2}.  It should be emphasized that  all of these solitary waves have a peaked crest.   These illustrate that   the generalized equation (\ref{geq:generalized:2}) also admits peaked solitary waves  for arbitrary constants $\alpha$ and $\beta$ even when $\omega \neq 0$, although the corresponding equation is {\em not}  integrable in general cases.   Therefore,  the existence of peaked solitary waves might be a common property of the generalized equation (\ref{geq:generalized:2}). 

As pointed out by  Constantin and Molinet \cite{Constantin2000-B},  all of these peaked solitary waves  should be understood mathematically as weak solutions of the generalized equation (\ref{geq:generalized:2}).   However,  physically,  this  kind of discontinuity of wave elevation widely appears in fluid mechanics, such as dam break \cite{Zoppou2000AMM}  in hydrodynamics and shock waves in aerodynamics, which have clear physical meanings.   Note also that  such kind of discontinuous problems belong to the so-called Riemann problem \cite{Bernetti2008JCP, Rosatti2010JCP, Wu2008IJNMF, Zoppou2000AMM},  a classical field of fluid mechanics.    

\section{Concluding remarks}

In this paper,  using the HAM-based Mathematica package BVPh 1.0 as a useful tool,  we illustrate that, like the CH equation (\ref{geq:CH}),  the DP equation (\ref{geq:DP}) also admits peaked solitary waves even when $\omega \neq 0$.  Furthermore,  we illustrate in a similar way that the generalized equations (\ref{geq:generalized:1}) and (\ref{geq:generalized:2})  also admit peaked solitary waves even when $\omega \neq 0$, and  the corresponding equation is {\em not}  integrable!     As pointed out by Liao \cite{Liao-arXiv-KdV},  nearly all mainstream models of shallow water waves  admit peaked solitary waves.    This work  further confirms  this view-point.     This is mainly because the fully nonlinear water wave equations  might  admit  peaked solitary waves, as pointed out by Liao \cite{Liao-arXiv-SW}.   Therefore,  mathematically speaking,   peaked solitary waves widely exist and might be a common property of models  for  water  waves,  although it should be carefully investigated whether or not such kind of peaked solitary waves  indeed  could be regarded as  weak  solutions  and  besides have physical meanings.        

\section*{Acknowledgement}

This work is partly supported by the State Key Lab of Ocean Engineering (Approval No. GKZD010056-6) and the National Natural Science Foundation of China.

\pagebreak\newpage

\bibliographystyle{unsrt}

\pagebreak\newpage

\begin{center}
{\bf Appendix\\ 
\vspace{0.5cm} 
 The use of HAM-based Mathematica package BVPh 1.0}
\end{center}

The  Mathematica package BVPh 1.0 for nonlinear boundary-value/eigenvalue problems is developed and issued by Liao  \cite{LiaoBook2012}  (Part II), which is based on the HAM and is free available online.   Using the BVPh 1.0, it is straightforward to gain the analytic approximations of the peaked solitary waves of  Eqs.  (\ref{geq:generalized:3}) and (\ref{bc:generalized:3}) for given $w(0)=A$.  Here, we briefly describe how to do it in case of $c = 1, \omega = 1/4$, $\alpha = 4, \beta = 3$  and $A=1/10$, corresponding to the DP equation (\ref{geq:DP}).

\begin{enumerate}
  \item First, download the  BVPh 1.0 (the code file is named by BVPh\_1.0.txt) online ( http://numericaltank.sjtu.edu.cn/BVPh.htm )
       and save it in a directory such as C:/math/DP as an example.

  \item Then, run the computer algebra system Mathematica, and type the following command one by one:
\end{enumerate}
{\small
\begin{verbatim}
SetDirectory["C:\math\DP"];

<<InputDP.txt
\end{verbatim}
}

The file named InputDP.txt contains the following Mathematica commands and necessary definitions for BVPh 1.0:
{\small
\begin{verbatim}
(* Install the BVPh 1.0  *)
<<BVPh1_0.txt;

(* Define the physical and control parameters *)
TypeEQ      =  1;
ApproxQ     =  0;
ErrReq      =  10^(-30);  
zRintegral  =  10;


(* Define the governing equation *)
ALPHA = 4 ;
BETA  = 3;
mu2 = 1-2*omega/c;
f[z_,u_,lambda_] := D[u,{z,3}]-mu2*D[u,z]  \
    + ALPHA*u*D[u,z] - BETA*D[u,z]*D[u,{z,2}] - u*D[u,{z,3}] ;
If[BETA == 2 && ALPHA == 3, Print["CH equation"]];
If[BETA == 3 && ALPHA == 4, Print["DP equation"]];


 
(* Define Boundary conditions *)
zR  = Infinity;
OrderEQ  = 3;
BC[1,z_,u_,lambda_] := Limit[u-A, z -> 0 ];
BC[2,z_,u_,lambda_] := Limit[u, z -> zR ];
BC[3,z_,u_,lambda_] := Limit[D[u,z], z -> zR ];

(* Define initial guess *)
mu = Sqrt[mu2];
u[0]  = A*Exp[-mu*z];

(* Define output term *)
output[z_,u_,k_]:= Print["output = ",D[u[k],z] /. z->0//N];

(* Defines the auxiliary linear operator *)
L[u_] := D[u,{z,3}] - mu2 * D[u,z];

(*  Print input and control parameters  *)
PrintInput[u[z]];

(* Set convergence-control parameter c0 and physical parameters *)
c0    =  -1 ;
A     =  1/10;
omega =  1/4;
c     =   1;
Print[" c0  =  ",c0, "  omega  =  ",omega, "  c   = ", c, "  A  = ",A];


(*  Gain up to 10th-order HAM approxiamtion *)
BVPh[1,10];

(*  Get results in the whole domain  *)
For[k=0,k<=10,k++,W[k] = U[k] /. z-> Abs[x]];

(* Show the 5th and 10th-order approximation  *)
Plot[{W[5],W[10]},{x,-10,10},PlotRange->{Min[A,0],Max[A,0]}]

\end{verbatim}
}

\end{document}